\documentclass{elsart}
\usepackage{graphicx}
\usepackage{dcolumn}
\usepackage{amsmath}
\usepackage{amssymb}

\begin{document}
\begin{frontmatter}
\title{How a Long Bubble Shrinks: a Numerical Method for an Unforced
Hele-Shaw Flow}
\author{Arkady Vilenkin and  Baruch Meerson}
\address{The Racah Institute  of  Physics, The Hebrew University
of Jerusalem, Jerusalem 91904, Israel}
\begin{abstract}
We develop a numerical method for solving a free boundary problem
which describes shape relaxation, by surface tension, of a long and
thin bubble of an inviscid fluid trapped inside a viscous fluid in a
Hele-Shaw cell. The method of solution of the exterior Dirichlet
problem employs a classical boundary integral formulation. Our
version of the numerical method is especially advantageous for
following the dynamics of a very long and thin bubble, for which an
asymptotic scaling theory has been recently developed. Because of
the very large aspect ratio of the bubble, a direct implementation
of the boundary integral algorithm would be impractical. We modify
the algorithm by introducing a new approximation of the integrals
which appear in the Fredholm integral equation and in the integral
expression for the normal derivative of the pressure at the bubble
interface. The new approximation allows one to considerably reduce
the number of nodes at the almost flat part of the bubble interface,
while keeping a good accuracy. An additional benefit from the new
approximation is in that it eliminates numerical divergence of the
integral for the tangential derivative of the harmonic conjugate.
The interface's position is advanced in time by using explicit node
tracking, whereas the larger node spacing enables one to use larger
time steps. The algorithm is tested on two model problems, for which
approximate analytical solutions are available.
\end{abstract}

%\PACS 02.70.Pt 47.11.+j 47.15.Gf 47.15.Hg
\end{frontmatter}
\maketitle

\section{Introduction}
Let a bubble of low-viscosity fluid (say, air) get trapped inside a
high-viscosity fluid (say, oil) in a quasi-two-dimensional Hele-Shaw
cell: two parallel plates with a narrow gap between them. What will
happen to the shape of the bubble, if the (horizontal) plates of the
Hele-Shaw cell are perfectly smooth, and the two fluids are
immiscible? The answer depends on the initial bubble shape. A
perfectly circular bubble (or an infinite straight strip) will not
change, while a bubble of any other shape will undergo
surface-tension-driven relaxation until it either becomes a perfect
circle, or breaks into two or more bubbles, which then become
perfect circles. The bubble shape relaxation process is non-local,
as it is mediated by a flow in the external viscous fluid. The
two-dimensional surface-tension-driven flow can be called an
unforced Hele-Shaw (UHS) flow. This is in contrast to forced
Hele-Shaw flows that have been in the focus of hydrodynamics and
nonlinear and computational physics for the last two decades
\cite{Langer1,BKD,Kessler,Casademunt1,Casademunt2}. In rescaled
variables, the UHS flow is described by the solution of the
following free boundary problem, see \textit{e.g.} Refs.
\cite{CLM,VMS}:
\begin{equation} \label{LPLC}
\nabla^{2}p(q)=0\qquad \mbox{for}\qquad q\in E,
\end{equation}
\begin{equation} \label{GT}
p(q)=\mathcal{K}\qquad \mbox{for}\qquad q\in \gamma,
\end{equation}
%{\cal K}
\begin{equation} \label{VN}
v_{n}(q)=-\nabla_{n}p(q)\qquad \mbox{for}\qquad q\in \gamma,
\end{equation}
where $E$ is an unbounded region of the plane, external to the
bubble interface $\gamma$, $v_{n}$ is the normal velocity of the
interface, the index $\textsc{n}$ denotes the component of vectors
normal
to the interface, and $K$ is the local curvature of the %{\cal K}
interface. The pressure $p$ is bounded at infinity. The free
boundary problem (\ref{LPLC})-(\ref{VN}) splits into two
sub-problems:
\begin{enumerate}
\item{Solving the exterior Dirichlet problem (\ref{LPLC}) and
(\ref{GT}) and  calculating $v_{n}(q)$ from Eq. (\ref{VN}).}
\item{Advancing the interface $\gamma$ in time with the known
$v_{n}(q)$.}
\end{enumerate}

The free boundary problem (\ref{LPLC})-(\ref{VN}) represents an
important example of area-preserving curve-shortening motion
\cite{ConstantinPugh}, but it is not integrable. Moreover, the only
analytical solution to this problem, available until recently, was
the approximate solution following from a linear stability analysis
of a slightly deformed circular or flat interface \cite{linear}.
Recently an asymptotic scaling theory has been developed for a
non-trivial case when the inviscid fluid occupies, at $t=0$, a
half-infinite (or, physically, very long) strip \cite{VMS}. It
turned out that this somewhat unusual initial condition provides a
useful characterization of the UHS flow, as the evolving strip,
which develops a dumbbell shape, exhibits approximate
self-similarity with non-trivial dynamic exponents \cite{VMS}.
Predictions of the scaling analysis have been verified numerically
in Ref. \cite{VMS} by using a boundary integral algorithm, tailored
to the very large aspect ratio of the bubble. The present paper
describes this algorithm in detail.

A multitude of numerical methods have been suggested in the recent
years for simulating different variants of Hele-Shaw flows. Boundary
integral methods, which deal directly with the interface between the
two fluids, are advantageous compared to methods of finite elements
and finite differences. Methods based on conformal mapping
techniques have long been used in this class of problems (see,
\textit{e.g.} Refs. \cite{BKD,DKZ,Tanv}). However, they apply most
naturally to the case of \textit{zero} surface tension and are less
convenient when surface tension is non-zero \cite{HLSH}.  Still
another numerical strategy is phase field methods. Folch \textit{et
al.} \cite{Folch1,Folch2} developed a phase field method for an
arbitrary ratio of the viscosities of the two fluids. Unfortunately,
their method becomes inefficient when the viscosity contrast is too
high \cite{Folch2}. To remind the reader, the viscosity contrast is
infinite in the case under consideration in the present work.
Glasner \cite{Glasner} developed a phase field method for a
description of a bubble of a high-viscosity fluid trapped in a
low-viscosity fluid. We are unaware of any phase-field approach
which would deal with the opposite case, which is under
investigation in the present work: a low-viscosity bubble in a
high-viscosity fluid.

The present work suggests a numerical algorithm for solving the free
boundary problem (\ref{LPLC})-(\ref{VN}) in the special case of a
very long bubble. It is well-known (but still remarkable) that the
exterior Dirichlet problem (\ref{LPLC}) and (\ref{GT}) can be
formulated in terms of a Fredholm integral equation of the second
kind for an effective density of the dipole moment \cite{POTTH}. A
na\"{\i}ve formulation, however, would lead to non-existence of
solution by the Fredholm alternative \cite{GGM}. To overcome this
difficulty, Greenbaum \textit{et al.} \cite{GGM} implemented in
their algorithm a modification of the Fredholm equation due to
Mikhlin \cite{Mikhlin}. The modified Fredholm equation has a unique
solution for any smooth $\gamma$ and integrable $\mathcal{K}$
\cite{Mikhlin}. Greenbaum \textit{et al.} developed an efficient
numerical algorithm (which is also valid for multiple bubbles) by
discretization. However, the geometry of a very long and thin
bubble, that we are mostly interested in, defines widely different
length scales in the problem. Rapid variations of the dipole moment
density at the highly curved ends of the bubble naturally
necessitate a small spacing between the interface nodes. It is less
natural, however, that, in a straightforward approach, one must keep
the node spacing much smaller than the bubble thickness \textit{over
the whole bubble interface}. Indeed, as we show below, the typical
length scale of the variation of the kernel of the integral equation
is comparable to the bubble \textit{thickness} which, during the
most interesting part of the long bubble dynamics, remains almost
unchanged.  Apart from being computationally inefficient, the
straightforward approach would cause a problem for explicit tracking
of nodes, as the stability criterion, intrinsic in the explicit
method, demands a time step less then a constant $\mathcal{O}(1)$
multiplied by the node spacing cubed \cite{Beale}. In this work we
turned this obstacle into advantage, by employing the fact that the
length scale of variation of the solution, over the most of the
bubble interface, is much greater than the bubble thickness. We
constructed a new approximation of the integral entering the
Fredholm equation, by representing the sought dipole moment density
as a piecewise constant function, and the bubble interface shape
function as a piecewise linear function. As a result, the integral
is approximated by a sum, each term of which is equal to a local
value of the dipole moment density multiplied by an integral of the
kernel between two neighboring nodes. Fortunately, the latter
integral can be calculated analytically. The new approximation
allowed us to considerably increase the node spacing over the most
of the bubble interface, while keeping a good accuracy.

Having found an approximated solution $p$ in the form of a double
layer potential, one needs to compute the normal derivative of the
solution $\nabla_{n}p(q\in \gamma)$. In a straightforward
realization of the boundary integral formulation this would result
in a hypersingular integral, see Ref. \cite{GGM}. To overcome this
difficulty, one resorts to theory of analytic functions and computes
the harmonic conjugate $V(q)$. By virtue of the Cauchy-Riemann
equations, the tangential derivative of $V(q)$ is equal to the
desired normal derivative of $p$.  The harmonic conjugate $V(q)$ has
the form of a principal value integral, over the interface, of the
dipole moment density multiplied by a kernel, which is a function of
coordinates of two points, $q$ and $g$, belonging to the interface.
This kernel diverges when the integration variable $g$ coincides
with $q$. Here we again employ the large scale difference at the
flat part of the bubble and use the same approximation as in the
Fredholm equation. As an additional benefit, the numerical
divergence of the integrand of the harmonic conjugate $V$ is
avoided. As a result, we do not need to use even nodes to compute
$V$ at odd nodes and vice-versa, as suggested in Ref. \cite{GGM}.

Here is a layout of the rest of the paper. Section 2 deals with the
numerical solution of the exterior Dirichlet problem, and with the
computation of the normal derivative of the solution at the
interface. We briefly review the boundary integral method for an
exterior Dirichlet problem and motivate the need for its
modification for very long bubbles. Then we formulate our discrete
approximation. In Section 3 we briefly describe a simple explicit
integration which we used to track the bubble interface. Section 4
presents the results of code testing, while Section 5 presents the
Conclusions.

\section{Exterior Dirichlet Problem}
\subsection{Boundary integral formulation}
\begin{figure}%[ptb]
\centerline{\includegraphics[width=9.5cm,clip=] {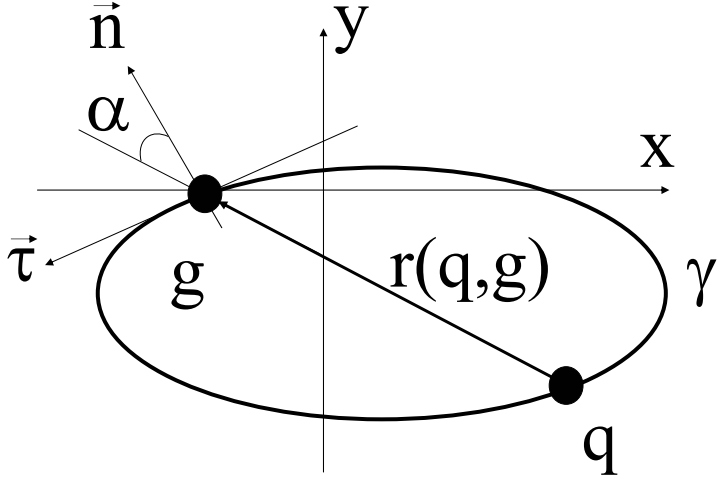}}
\caption{Geometry of the kernel $K(q,g)$. $\vec{\tau}$ and $\vec{n}$
are the tangential and outward normal directions, respectively.}
\label{fig1}
\end{figure}

Following Mikhlin \cite {Mikhlin}, we seek the solution $p(q)$ of
the problem(\ref{LPLC}) and (\ref{GT}) for a simply connected bubble
in a double layer potential representation:
\begin{equation} \label{PofQ}
p(q)=\frac{1}{2\pi}\oint_{\gamma}\left[1+K(q,g)\right] \mu(g)\,
dg\;\;\; \mbox{for}\;\;q\in E,
\end{equation}
where $\mu(g)$ is an unknown dipole density at the point $g$ of the
interface, and $dg$ is the element of arclength including the point
$g$. The kernel $K(q,g)=\cos \alpha/|\vec{r}(q,g)|$ follows from
classical potential theory \cite{POTTH}. Here $\alpha$ is the angle
between the outward normal to the interface at the point $g$ and the
vector $\vec{r}(q,g)$, see Fig.~\ref{fig1}.

The boundary condition (\ref{GT}) can be rewritten as an integral
equation for $\mu(q)$, see, \textit{e.g.}, \cite{POTTH}:
\begin{equation} \label{INTEQ1}
-\mu(q) + \frac{1}{\pi}\oint_{\gamma}\left[1+K(q,g)\right] \mu(g)\,
dg = 2\mathcal{K}(q)\,.
\end{equation}
%{\cal K}
%The solution, described by Eq. (\ref{PofQ}), is discontinuous at the
%interface $\gamma$, and Eq. (\ref{INTEQ1}) satisfies the jump
%relation \cite{GGM,POTTH}.
That is, to compute $p(q)$  in Eq. (\ref{PofQ}), one needs to solve
the integral equation (\ref{INTEQ1}).  Mikhlin \cite{Mikhlin} showed
that Eq. (\ref{INTEQ1}) has a unique solution for any smooth
$\gamma$ and integrable $\mathcal{K}$, while $p(q)$ from Eq.
(\ref{PofQ}) is a harmonic function in the exterior, satisfying the
boundary condition Eq. (\ref{GT}). This representation was employed
by Greenbaum \textit{et al.} \cite{GGM} for numerical analysis.
%\cal K

For the purposes of the free boundary problem
(\ref{LPLC})-(\ref{VN}) one only needs the value of
$\nabla_{n}p(q)$, $p\in \gamma$. A straightforward calculation of
$\nabla_{n}$ from the double layer potential would yield a
hypersingular integral, see below. One circumvents this difficulty
by resorting to theory of analytic functions, see Ref. \cite{GGM}
and references therein. Suppose $\mu(q)$ is known and introduce the
quantity
$$\tilde{p}(q)=\frac{1}{2\pi}\oint_{\gamma}\mu(g)K(q,g)\,dg\,,$$
which differs from $p(q)$ only by a constant, as
%=p(q)-\mbox{const},$$ since
$\oint_{\gamma}\mu(g)\,dg=\mbox{const}.$ Obviously,
$\nabla_{n}\tilde{p}=\nabla_{n}p$.  It is known \cite{GGM} that
$\tilde{p}(q)$ is the real part of the Cauchy integral
$$\frac{1}{2\pi i}\oint_{\gamma}\frac{\mu(\zeta)}{\zeta-z}d\zeta=\tilde{p}(z)+iV(z),$$
where we have identified the points $q$ and $g$ on the plane with
respective complex numbers $z$ and $\zeta$. Then $\tilde{p}$ and its
harmonic conjugate $V$ satisfy the Cauchy-Riemann equations, so that
\begin{equation}\label{CR}
\tilde{p}_{n}=V_{\tau}\,,
\end{equation}
where the indices $n$ and $\tau$ stand for the normal and tangential
derivatives, respectively. The kernel $K(q,g)$ can be written as
follows:
$$K(q,g)\,dg=\frac{-(y_{g}-y_{q})\,dx_{g}+(x_{g}-x_{q})\,dy_{g}}{r^{2}(q,g)}\,,$$
where $x$ and $y$ are the Cartesian coordinates of the respective
points. After a simple algebra we obtain
\begin{equation}\label{Vofz}
V(q)=-\frac{1}{2\pi}\oint_{\gamma}\frac{\mu(x_{g},y_{g})(x_{g}-x_{q})}{r^{2}(q,g)}\,dx_{g}+
\frac{\mu(x_{g},y_{g})(y_{g}-y_{q})}{r^{2}(q,g)}\,dy_{g}\,.
\end{equation}

\subsection{Discrete approximation}

Let us parameterize the closed interface $\gamma$ of the bubble:
$x=x(\sigma)$, $y=y(\sigma)$, $0 \leq \sigma \leq M$, $x(0)=x(M)$,
$y(0)=y(M)$, where $x$ and $y$ are the Cartesian coordinates of a
point belonging to the interface. In the parametric form Eq.
(\ref{INTEQ1}) becomes
\begin{equation} \label{INTEQPR}
- \mu(\sigma) +
\frac{1}{\pi}\int_{0}^{M}\mu(\xi)\left[1+\kappa(\sigma,\xi)
\right]\sqrt{\dot{x}^{2}+\dot{y}^{2}} \,d\xi  =
2\mathcal{K}(\sigma)\,,
\end{equation}
%{\cal K}
where
\begin{equation}\label{Kksi}
\kappa(\sigma,\xi)=\frac{\dot{y}[x(\xi)-x(\sigma)]-\dot{x}[y(\xi)-y(\sigma)]}
{\{[x(\sigma)-x(\xi)]^{2}+[y(\sigma)-y(\xi)]^{2}\}\sqrt{\dot{x}^{2}+\dot{y}^{2}}}\,,
\end{equation}
while $\dot{x}=dx/d\xi$ and $\dot{y}=dy/d\xi$. The harmonic
conjugate takes the form
\begin{equation}\label{Vofksi}
V(\sigma)=-\frac{1}{2\pi}\int_{0}^{M}\mu(\xi)
\frac{\dot{x}[(x(\xi)-x(\sigma)]+\dot{y}[(y(\xi)-y(\sigma)]}{[(x(\xi)-x(\sigma)]^{2}+
[(y(\xi)-y(\sigma)]^{2}}\,d\xi.
\end{equation}
Note that the kernel $\kappa$ is continuous as $\xi \to \sigma$. On
the contrary, the integrand in  the last expression diverges as $\xi
\to \sigma$, and the integral exists only as a principal value.
\begin{figure}%[ptb]
\centerline{\includegraphics[width=8.5cm,clip=] {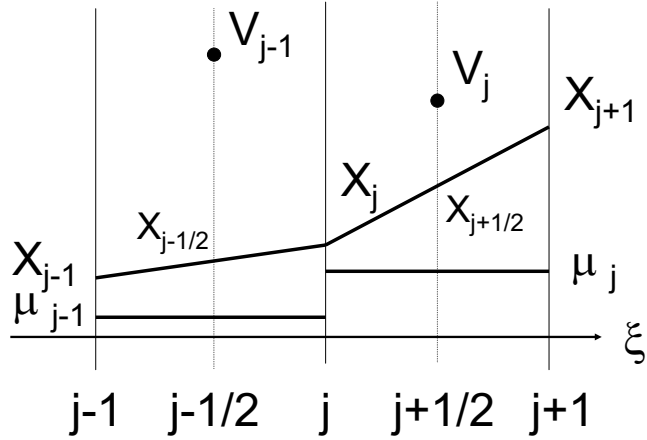}}
\caption{The discrete approximation scheme. Here $\xi_{j}=j.$}
\label{fig2}
\end{figure}
In the main case of our interest the bubble length is much greater
than its thickness $\Delta$. In the almost flat parts of the
interface $\dot{y} \approx 0$. Now, $y(\sigma)-y(\xi)\sim \Delta$
when the points $\sigma$ and $\xi$ belong to the different (upper
and lower) parts of the interface, while $y(\sigma)-y(\xi)\approx 0$
when they belong to the same part of the interface. Then, using the
relation $\dot{x}=dx/d\xi$, we can estimate the kernel $\kappa$ as
\begin{equation}\label{kappadelta}
\kappa(\sigma,\xi)\,d\xi\approx\frac{-\Delta}{[x(\sigma)-x(\xi)]^{2}+\Delta^{2}}\,dx\,,
\end{equation}
when $\sigma$ and $\xi$ belongs to the different parts of the
interface, while $\kappa \approx 0$ when they belong to the same
part. Equation (\ref{kappadelta}) shows that the typical scale of
variation of the kernel (\ref{Kksi}) over the almost flat part of
the interface is of order of the bubble thickness $\Delta$. A
similar estimate applies to the fraction entering the integrand of
Eq. (\ref{Vofksi}). A straightforward discretization would then
require a node spacing much less than $\Delta$. Instead, we rewrite
Eq. (\ref{INTEQPR}) as
%{\cal K}
\begin{equation}\label{inteqsm}
-\mu(\sigma) +
\frac{1}{\pi}\sum_{j=0}^{m-1}\int_{\xi_{j}}^{\xi_{j+1}}\mu(\xi)[1+\kappa(\sigma,\xi)
]\sqrt{\dot{x}^{2}+\dot{y}^{2}}\,d\xi  = 2\mathcal{K}(\sigma)\,,
\end{equation} where
$\xi_{0}=0$, $\xi_{m}=M$, and $\xi_{j+1}>\xi_{j}$, $j=0, 1, 2,
\dots, m-1$. Introduce a piecewise linear approximation for $x(\xi)$
and $y(\xi)$ (see Fig. \ref{fig2}):
\begin{equation}\label{xappr}
x(\xi)=k_{j}^{x}\xi+b_{j}^{x}, \quad y(\xi)=k_{j}^{y}\xi+b_{j}^{y}, %\quad \xi_{j}\leq\xi\leq\xi_{j+1},\quad j=0, 1, 2, \dots, m-1,
\end{equation}
where $\xi_{j}\leq\xi\leq\xi_{j+1},\quad j=0, 1, 2, \dots, m-1,$
 $k_{j}^{x}=(x_{j+1}-x_{j})/(\xi_{j+1}-\xi_{j})$,
$b_{j}^{x}=(\xi_{j+1}x_{j}-\xi_{j}x_{j+1})/(\xi_{j+1}-\xi_{j})$,
$k_{j}^{y}=(y_{j+1}-y_{j})/(\xi_{j+1}-\xi_{j})$,
$b_{j}^{y}=(\xi_{j+1}y_{j}-\xi_{j}y_{j+1})/(\xi_{j+1}-\xi_{j})$,
$x_{j}=x(\xi_{j})$, $y_{j}=y(\xi_{j}),$ and a piecewise constant
approximation for $\mu$:
\begin{equation}\label{muappr}
\mu(\xi)=\mu_{j}=\mbox{const}, \qquad \xi_{j}\leq\xi\leq\xi_{j+1},
\end{equation}
Note that $\dot{x}(\xi_{j}\leq\xi\leq\xi_{j+1})=k^{x}_{j},$
$\dot{y}(\xi_{j}\leq\xi\leq\xi_{j+1})=k^{y}_{j}$. The kernel
(\ref{Kksi}) is therefore approximated as
$$\kappa(\sigma,\xi)=\frac{k^{y}_{j}[k^{x}_{j}\xi+b^{x}_{j}-x(\sigma)]-
k^{x}_{j}[k^{y}_{j}\xi+b^{y}_{j}-y(\sigma)]}
{\{[k^{x}_{j}\xi+b^{x}_{j}-x(\sigma)]^{2}+[k^{y}_{j}\xi+b^{y}_{j}-y(\sigma)]^{2}\}S_{j}}
%=$$$$
=\frac{Q_{j}(\sigma)/S_{j}}
{S_{j}^{2}\xi^{2}+B_{j}(\sigma)\xi+C_{j}(\sigma)},$$
where
$$
S_{j}=\sqrt{(k^{x}_{j})^{2}+(k^{y}_{j})^{2}}, \quad
Q_{j}(\sigma)=k^{y}_{j}[b^{x}_{j}-x(\sigma)]-k^{x}_{j}[b^{y}_{j}-y(\sigma)]\,,
$$
$$
B_{j}(\sigma)=2\{k^{x}_{j}[b^{x}_{j}-x(\sigma)]+k^{y}_{j}[b^{y}_{j}-y(\sigma)]\},\;\;\mbox{and}\;\;
C_{j}(\sigma)=[b^{x}_{j}-x(\sigma)]^{2}+[b^{y}_{j}-y(\sigma)]^{2}\,.
$$
The integrals in (\ref{inteqsm}) can be calculated analytically:
$$\int_{\xi_{j}}^{\xi_{j+1}}\mu(\xi)[1+\kappa(\sigma,\xi)
]\sqrt{\dot{x}^{2}+\dot{y}^{2}}\,d\xi
%=$$$$
=\mu_{j}\int_{\xi_{j}}^{\xi_{j+1}}\left[S_{j}+\frac{Q_{j}(\sigma)}
{S_{j}^{2}\xi^{2}+B_{j}(\sigma)\xi+C_{j}(\sigma)}\right]\, d\xi = $$
$$=\mu_{j}\left\{S_{j}(\xi_{j+1}-\xi_{j})
%+$$$$
+\frac{1}{Q_{j}(\sigma)}
\left[\arctan\frac{2S_{j}^{2}\xi_{j+1}+B_{j}(\sigma)}{2Q_{j}(\sigma)}-
\arctan\frac{2S_{j}^{2}\xi_{j}+B_{j}(\sigma)}{2Q_{j}(\sigma)}\right]\right\}.$$
It is convenient to define $\xi_{j}=j$, then $\xi_{j+1}-\xi_{j}=1$.
In our discretization scheme
$x(\sigma)=x_{i+1/2}=k_{i}^{x}(i+1/2)+b_{i}^{x}$ and
$y(\sigma)=y_{i+1/2}=k_{i}^{y}(i+1/2)+b_{i}^{y}$. Let us denote
$Q_{j}(\sigma)=Q_{ij}$, $B_{j}(\sigma)=B_{ij}$, and
$C_{j}(\sigma)=C_{ij}.$ The integrals in Eq. (\ref{inteqsm}) are
$$\int_{\xi_{j}}^{\xi_{j+1}}\mu(\xi)[1+\kappa(\sigma,\xi)
]\sqrt{\dot{x}^{2}+\dot{y}^{2}}\,d\xi=\pi \mu_{j}\textsc{A}_{ij},$$
where
%{\cal A}{\cal A}
$$\textsc{A}_{ij}=\frac{1}{\pi}\left\{S_{j}
%+$$$$
+\frac{1}{Q_{ij}}\arctan\frac{4S_{j}^{2}Q_{ij}}
{4Q_{ij}+(2S_{j}^{2}j+B_{ij})[2S_{j}^{2}(1+j)+B_{ij}]}\right\}.$$ We
have arrived at a set of linear algebraic equations with respect to
$\mu_{j}$, which is our approximation of the integral equation
(\ref{INTEQ1}):
\begin{equation} \label{INTEQdscr}
\sum_{j=0}^{m-1}(\textsc{A}_{ij}-\delta_{ij})\mu_{j}=2\mathcal{K}_{i},\quad
i=0, 1, 2, 3, \dots, m-1\,.
\end{equation}
%{\cal A}{\cal K}
%{\cal K}
We approximate the interface curvature $\mathcal{K}(\sigma)$ by
finite differences:
%{\cal K}{\cal K}
$$\mathcal{K}(\sigma=i+1/2)=\mathcal{K}_{i}=\frac{\ddot{y}_{i+1/2}\dot{x}_{i+1/2}-\ddot{x}_{i+1/2}\dot{y}_{i+1/2}}
{[(\dot{y}_{i+1/2})^{2}+(\dot{x}_{i+1/2})^{2}]^{3/2}},$$ where
$\dot{x}_{i+1/2}=k_{i}^{x}$, $\dot{y}_{i+1/2}=k_{i}^{y}$,
$\ddot{x}_{i+1/2}=x_{i+2}-x_{i+1}-x_{i}+x_{i-1}$, and
$\ddot{y}_{i+1/2}=y_{i+2}-y_{i+1}-y_{i}+y_{i-1}$.
%{\cal A}
Importantly, our approximation scheme yields the principal value of
the integral (\ref{Vofksi}) automatically. Furthermore, we can
directly compute the coefficients $\textsc{A}_{ij}$, using the same
expression for $i\neq j$ and $i=j$, where the kernel (\ref{Kksi})
has a removable discontinuity. The method suggested in \cite{GGM}
prescribes instead to use an analytic evaluation of the kernel at
the point of removable discontinuity.

We solved the algebraic equations  (\ref{INTEQdscr})  by  an
iterative refinement method after a LU factorization of  the matrix.
As the maximum number of equations in the examples that we
considered (see below) did not exceed $1100$, there was no need to
use more sophisticated methods.

\subsection{Grid}
Most of our results were obtained with the version of the code which
assumed a four-fold symmetry of the bubble. This allowed us to work
with a one quarter of the interface and reduce the number of nodes
by 4. In the beginning of the bubble relaxation, the solution varies
rapidly in the region of the lobes, and very slowly in the flat
region of the bubble. Therefore one should employ here a non-uniform
grid. At later times, when the aspect ratio of the bubble becomes
comparable to unity, the code switches to a uniform grid. For the
non-uniform grid we used an exponential spacing. Here the node
spacing grows exponentially from the lobe's end to the middle of the
flat part of the bubble. To  generate the node distribution we use
the following procedure. Let the quarter of the interface perimeter
be $\Pi$, the specified number of nodes be $m$, and the specified
\textit{smallest} spacing in the lobe region be $h_{0}$.

If $\Pi>h_{0}(m-1)$, the exponential grid is used. Here we introduce
the quantity $\eta$ which satisfies the condition
\begin{equation}\label{eta}
\Pi=h_{0}+\eta h_{0}+\eta^{2} h_{0} + \ldots + \eta^{m-2} h_{0}=
\frac{h_{0}(1-\eta^{m-1})}{1-\eta}\,,
\end{equation}
solve Eq. (\ref{eta}) numerically  for $\eta$, use a discrete
arclength parametrization: $\xi_{1}=0,\quad
\xi_{2}=h_{0},\dots,\xi_{k}=\eta^{k-2}
h_{0},\dots,\xi_{m}=\eta^{m-2} h_{0}$, and calculate the arrays
$x=x(\xi_{i})$ and $y=y(\xi_{i})$, where $i=1, 2, \dots, m$.

In the process of the interface evolution $\Pi$ decreases with time,
so one can reduce the number of the grid nodes. Furthermore, as the
nodes in our code move like lagrangian particles (see Section 3),
the node spacing in the lobe region decreases with time even faster.
If left unattended, this would cause instability of the node
tracking (see Section 3), as the maximum allowed time step is
proportional to the node spacing cubed \cite{Beale}.   Therefore,
when the minimum node spacing decreases below $\xi_{2}=0.8 h_{0}$,
we redistribute the nodes: we look for the new value of $\eta$,
corresponding to the updated value of $\Pi$, calculate the new array
of $\xi$, and determine the new arrays $x$ and $y$ by linear
interpolation.

When the perimeter goes down so that $\Pi\leq h_{0}(m-1)$, we switch
to a uniform grid. Here we calculate a new $m$: $m=[\Pi/h_{0}]+1$,
where $[a]$ is an integer number such that $0\leq a-[a]<1$, and fine
tune $h_0$ so that $h_{0}=\Pi/(m-1)$.

Finally, the choice of $h_0$ is dictated by a compromise between the
desired accuracy and the value of $m$ which determines the size of
the matrix $A_{ij}$.

\subsection{Calculation of the normal velocity}
After the set of linear equations (\ref{INTEQdscr}) is solved, and
the quantities $\mu_{i}$ are found, we compute the harmonic
conjugate $V$. The same approximation, applied to Eq.
(\ref{Vofksi}), yields:
$$V_{i}=-\frac{1}{2\pi}\sum_{j=0}^{m-1}\mu_{j}F_{ij}\,,$$
where
$$F_{ij}=\int_{j}^{j+1}
\frac{k_{j}^{x}(k_{j}^{x}\xi+b_{j}^{x}-x_{i+1/2})+k_{j}^{y}(k_{j}^{y}\xi+b_{j}^{y}-y_{i+1/2})}
{(k_{j}^{x}\xi+b_{j}^{x}-x_{i+1/2})^{2}+(k_{j}^{y}\xi+b_{j}^{y}-y_{i+1/2})^{2}}\,d\xi
$$
$$=\frac{1}{2}\int_{j}^{j+1}\frac{2S_{j}^{2}\xi+B_{ij}}
{S_{j}^{2}\xi^{2}+B_{ij}\xi+C_{ij}}\,d\xi
%=$$$$
=\frac{1}{2}\ln\frac{S_{j}^{2}(j+1)^{2}+B_{ij}(j+1)+C_{ij}}{S_{j}^{2}j^{2}+B_{ij}j+C_{ij}}\,,$$
where the quantities $S_{j}$, $B_{ij}$ and $C_{ij}$ were defined
earlier. Again, the integral is calculated analytically. The
resulting formula for $V_{i}$ is the following:
\begin{equation} \label{Vfinal}
V_{i}=-\frac{1}{4\pi}\sum_{j=0}^{m-1}\mu_{j}
\ln\frac{S_{j}^{2}(j+1)^{2}+B_{ij}(j+1)+C_{ij}}{S_{j}^{2}j^{2}+B_{ij}j+C_{ij}},
\end{equation}
where $V_{i}=V(\sigma=i+1/2),$ $i=0,1,2,3,\dots, m-1$, see Fig.
\ref{fig2}. Note that for $\xi=i+1/2$ the denominator of the
integrand in $F_{ij}$ vanishes, and the integrand diverges. To
overcome this problem, Ref. \cite{GGM} suggested to divide the mesh
into odd and even nodes and compute $V$ at the odd points by summing
over the even nodes, and vice-versa. Our analytical integration
yields the correct principal value of the integral, so there is no
need  to use the recipe of Ref. \cite{GGM}.

To compute the normal velocity of the interface we use the
Cauchy-Riemann equation (\ref{CR}) and approximate the derivative of
$V$ with respect to the arclength:
$$v_{n}(\sigma=i)=\frac{(V_{i+1}-V_{i})}
{\tilde{s}_{i}}\,,$$
where
$$\tilde{s}_{i}=\frac{1}{2}\left[\sqrt{(x_{i+1}-x_{i})^{2}+(y_{i+1}-y_{i})^{2}}+
%$$$$+
\sqrt{(x_{i}-x_{i-1})^{2}+(y_{i}-y_{i-1})^{2}}\right].$$
\section{Interface Tracking}
To track the interface, we use an explicit first-order integration:
$$x_{i}(t+\Delta t)=x_{i}(t)+\Delta t\, v_{n}(\sigma=i,t)\cos n_{i},$$
$$y_{i}(t+\Delta t)=y_{i}(t)+\Delta t\, v_{n}(\sigma=i,t)\sin n_{i},$$
where
$$\cos n_{i}=\frac{\dot{y}_{i}}{\sqrt{\dot{x}^{2}_{i}+\dot{y}^{2}_{i}}}\,, \quad
\sin
n_{i}=-\frac{\dot{x}_{i}}{\sqrt{\dot{x}^{2}_{i}+\dot{y}^{2}_{i}}}\,,$$
$\dot{x}_{i}=(\dot{x}_{i+1/2}+\dot{x}_{i-1/2})/2=(k^{x}_{i}+k^{x}_{i-1})/2$,
and $\dot{y}_{i}=(k^{y}_{i}+k^{y}_{i-1})/2$. We have assumed the
counter-clockwise direction of the interface parametrization, see
Fig. \ref{fig1}.

It is important to prescribe the time step $\Delta t$ properly. We
employ an ad-hoc criterion which demands that the node displacement
at each grid point be considerably less then the curvature radius
$R_i$ at that point: $\min |(\Delta t\, v_{n}(i))/R_{i}|\leq
\varepsilon,\quad 0\leq i \leq m-1.$ That is, we consider the
curvature radius $R_i$ as a natural local length scale of the
problem. A more convenient form of this criterion is
\begin{equation} \label{epsilon}
\Delta t=\varepsilon \min\{|R_{i}/v_{n}(i)|\}\,,
\end{equation}
where $\varepsilon$ is an input parameter which has to be
sufficiently small to satisfy the requirements of stability and
desired accuracy. In the exact formulation (\ref{LPLC})-(\ref{VN})
the bubble area must be constant in the process of relaxation. The
area conservation can be conveniently used for accuracy control of
the code.

\section{Numerical Results}
We present here some simulation results produced with our code for
two different sets of initial conditions. One of them describes the
decay of a small sinusoidal perturbation of a perfectly circular
bubble of inviscid fluid. An approximate analytical solution to this
problem is given by the linear stability analysis \cite{linear}, and
we used this solution to test the code.

The second initial condition describes a very long and thin strip of
inviscid fluid. In the process of its shrinking the bubble develops
a dumbbell shape, while the characteristic dimensions of the
dumbbell exhibit asymptotic scaling laws found in Ref. \cite{VMS}.

\subsection{Relaxation of a slightly perturbed circle}
\begin{figure}%[ptb]
\centerline{\includegraphics[width=7.5cm,clip=] {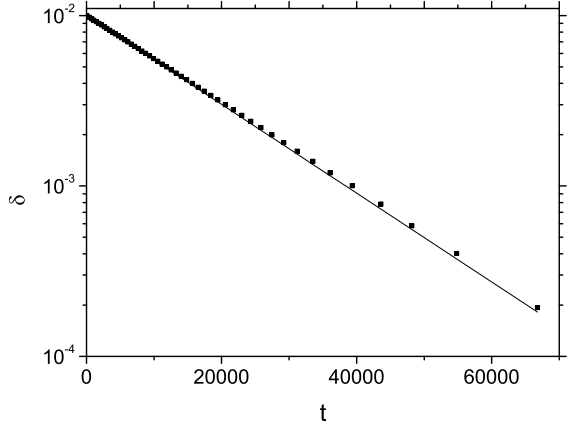}}
\caption{Shown in the logarithmic scale is the perturbation
amplitude $\delta$ as a function of time. The squares are the
simulation results, the solid line is the analytical prediction. }
\label{fig3}
\end{figure}

Let the initial shape of the interface be a circle with a small
sinusoidal perturbation:
$$
\rho(\varphi,0)=R_{0}[1+\delta(0)\sin(n\varphi)]\,,$$ where $\rho$
and $\varphi$ are the polar radius and angle, respectively, $R_{0}$
is the radius of the unperturbed interface, while $\delta(0)$ and
$n$ are the initial amplitude and azimuthal number of the
perturbation. The analytical solution provided by the linear theory
\cite{Paterson} is
$$
\rho(\varphi,t)=R_{0}[1+\delta(t)\sin(n\varphi)]\,,
$$
where the amplitude of the perturbation is
$$\delta(t)=\delta(0)\exp\left[-\frac{n(n^{2}-1)}{R_{0}^{3}}t\right],$$
A typical numerical result is presented in Fig. \ref{fig3}. The
parameters are $R_{0}=100,$ $\delta(0)=0.01$ and $n=4$. In the case
of $n=4$ the interface has a four-fold symmetry which allows a
direct application of our code. In this simulation the quarter of
the interface was described by 100 nodes. The initial spacing was
uniform. The code did not have to use the mesh interpolation in this
example. The parameter regulating the time step was
$\varepsilon=5\cdot10^{-5}$. As one can see, a very good agreement
with the analytical result is obtained.

\subsection{Relaxation of a long and thin bubble}
\begin{figure}%[ptb]
\includegraphics[width=13.5 cm,clip=] {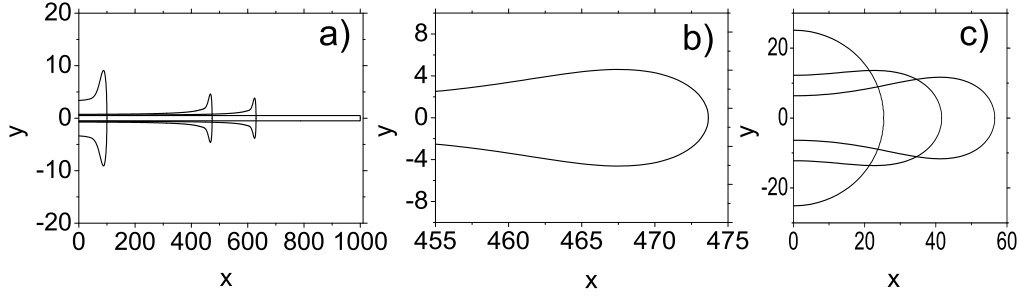} \caption{Figure a shows  a snapshot
of one half of the simulated system at $t=0$, $3670$, $7 020$, and
$24 840$. Notice the large difference between the horizontal and
vertical scales. Figure b shows the lobe of the dumbbell to scale at
$t=7 020$, while Figure c shows the computed bubble shape at late
times: $t=30 900$, $34 200$ and $48 000$.} \label{fig4}
\end{figure}

In the second setting the initial interface shape is a very long
rectangular strip. In the example we report here the initial strip
thickness was 1, and the initial length 2000. Here we could compare
the numerical results with the predictions of a recent asymptotic
scaling analysis \cite{VMS}. The interface shapes at different times
are presented in Fig. \ref{fig4}. It can be seen that the shrinking
strip acquires the shape of a dumbbell (or petal). At much later
times it approaches circular shape. By the end of the simulation (at
$t=48000$) the relative deviation of the observed shape from the
perfect circle, $[\rho_{max} (\varphi) - \rho_{min}
(\varphi)]/\rho_{min} (\varphi) \approx  0.013$.
\begin{figure}%[ptb]
\centerline{\includegraphics[width=11.0cm,clip=] {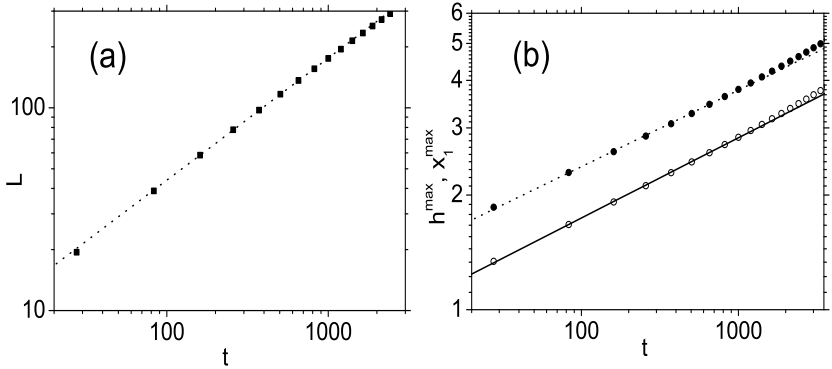}}
\caption{Figure a shows, in a log-log scale, the retreat distance
$L$ versus time and its power-law fit $2.75 \,t^{0.60}$. Figure b
shows, in a log-log scale, the maximum dumbbell height, $h^{max}$
(the empty circles), and the position of the maximum, $x_1^{max}$
(the filled circles), versus time, as well as their power-law fits
$0.66\,t^{0.21}$ and $0.94\,t^{0.20}$, respectively.}
\label{lengths}
\end{figure}

The asymptotic scaling analysis \cite{VMS} deals with the
intermediate stage of the relaxation. Introduce the retreat distance
$L(t)=1000-x_{tip}(t)$, where $x_{tip}(t)$ is the maximum abscissa
of all points belonging to the interface. One prediction of Ref.
\cite{VMS} is that, at intermediate times, $L(t)\propto t^{3/5}$.
Figure \ref{lengths}a shows a very good agreement of this prediction
with the simulation result. Additional predictions of asymptotic
scaling analysis deal with the time dependence of the maximum
dumbbell elevation $h^{max}(t)$, and of the abscissa of the
corresponding point of the interface $x^{max}(t)$. Let us introduce
a new variable: $x_{1}(x,t)=x_{tip}(t)-x$, the distance along the
$x$-axis between the tip of the dumbbell and a point $x$. In
particular, $x_{1}^{max}(t)=x_{tip}(t)-x^{max}(t)$. A comparison of
the simulation results with the predicted intermediate-time scaling
laws $h^{max}(t)\propto x_{1}^{max}(t) \propto t^{1/5}$ is shown in
Figure \ref{lengths}b, and again a very good agreement is observed.

\begin{figure}%[ptb]
\centerline{\includegraphics[width=8.0cm,clip=] {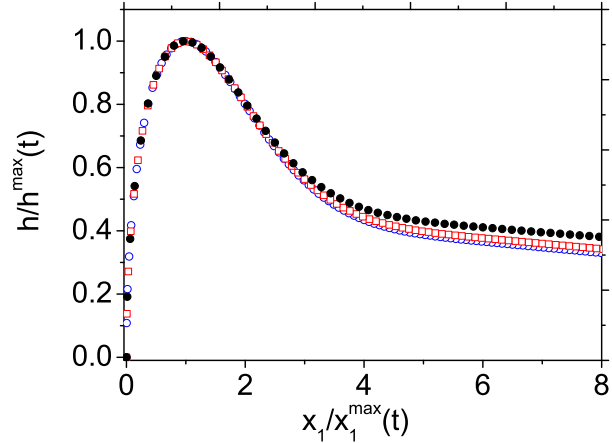}}
\caption{Self-similarity of the lobe. Shown is the shape function
$h(x_1,t)$, rescaled to the maximum dumbbell elevation, versus the
coordinate $x_1$, rescaled to the abscissa of the maximum, at times
$160.3$ (the filled circles), $1000$ (the squares), and $3010$ (the
empty circles).} \label{SS}
\end{figure}

To verify the self-similarity of the dumbbell shape in the lobe
region, predicted in Ref. \cite{VMS}, we introduce a new function
$h(x_1,t)$ so that $h[x_{1}(x,t),t]=y(x,t)$. Figure \ref{SS} shows
the spatial profiles of $h$ rescaled to the values of $h^{max}$
versus $x_1/x_1^{max}$ at three different times. The observed
collapse in the lobe region confirms the expected self-similarity.

The initial number of nodes in this simulation was 1100, and the
smallest spacing in the lobe region was 0.4.  With the grid
interpolation employed, the time-step parameter $\varepsilon=0.005$
proved sufficiently small to guarantee stability and good accuracy.
As the curvature of the interface goes down during the evolution,
the required time step increases significantly. It was  $1.7 \times
10^{-3}$ at $t=0$, $0.22$ at $t=3670$ and increased up to about 10
by the end of the simulation, at $t=48000$. We used the small
observed area loss of the bubble for accuracy control. The observed
area loss was less then $0.5\%$ for $t<10000$. By the end of the
simulation, at $t=48000$, the area loss reached only $2.8\%$.

\section{Conclusion}
We have developed and tested a new numerical version of the boundary
integral method for an exterior Dirichlet problem, which is
especially suitable for long and thin domains. The method allows one
to significantly reduce the number of the interfacial nodes. The new
method was successfully tested in a numerical investigation of the
shape relaxation, by surface tension, of a long and thin bubble,
filled with an inviscid fluid and immersed in a viscous fluid in a
Hele-Shaw cell. Here we confirmed the recent theoretical predictions
on the self-similarity and dynamic scaling behavior during an
intermediate stage of the bubble dynamics.

\section*{Acknowledgment}
This work was supported by the Israel Science Foundation, Grant No.
180/02.


\begin{thebibliography}{111}
\bibitem{linear} The damping rates of small sinusoidal
perturbations of circular and flat interfaces are given by the
zero-flow-rate limit of Eq.~(11) of Ref. \cite{Paterson} (for the
circular interface) and Eq.~(10) of Ref. \cite{ST} (for the flat
interface).
\end{thebibliography}
\end{document}